# First-principles modeling of zincblende AlN layer in Al-AlN-TiN multilayers

S. K. Yadav[1*], J. Wang[2], X.-Y. Liu[1*]

[1] Materials Science and Technology Division, MST-8, Los Alamos National Laboratory, Los Alamos, New Mexico 87545, USA

[2] Mechanical and Materials Engineering, University of Nebraska-Lincoln, Lincoln, NE 68588, USA

## Abstract

An unusual growth mechanism of metastable zincblende AlN thin film by diffusion of nitrogen atoms into Al lattice is established. Using first-principles density functional theory, we studied the possibility of thermodynamic stability of AlN as a zincblende phase due to epitaxial strains and interface effect, which fails to explain the formation of zincblende AlN. We then compared the formation energetics of rocksalt and zincblende AlN in fcc Al through direct diffusion of nitrogen atoms to Al octahedral and tetrahedral interstitials. The formation of zincblende AlN thin film is determined to be a kinetically driven process, not a thermodynamically driven process.

---

[*] Correspondence should be addressed to S. K. Yadav (syadav@lanl.gov) or X.-Y. Liu (xyliu@lanl.gov)

## 1. Introduction

Aluminum nitride (AlN) is a wide-gap optoelectronic material that is of considerable technology interest,[1] with many unique properties such as relatively high hardness, high thermal conductivity, etc. AlN thin films have also been grown on various substrates for application such as piezoelectric material.[2, 3] The ground state equilibrium structure of AlN at ambient temperature and pressure is B4 type, hexagonal wurtzite (*wz*-AlN). Additionally, AlN can also exist as a metastable B3 type, cubic zincblende structure (*zb*-AlN) or the high-pressure B1 type, cubic rock-salt variant (*rs*-AlN), as predicted using density functional theory (DFT) calculations.[4-6]

There has been a growing interest in the metastable cubic films of AlN recently, aiming to achieve novel and enhanced mechanical and functional properties that are not observed in the hexagonal structure. By tailoring substrates in terms of crystal structure, substrate orientation, and elastic mismatch, the cubic crystal structure of AlN layers can be grown in superlattice systems and the stability of the cubic structures strongly depends on their layer thickness. At an annealing temperature of 600 °C, epitaxial metastable *zb*-AlN was synthesized by the solid-state reaction between single crystal Al(001) and TiN(001).[7] In magnetron sputtering deposited AlN/TiN(001) epitaxial superlattices, the high-pressure *rs*-AlN was stabilized with AlN layer thickness less than or equal to 2.0 nm.[1] Using reactive sputtering, the epitaxial stabilization of *rs*-AlN was also observed in AlN/VN(001) and AlN/TiN(001) superlattices with critical layer thickness of AlN before transforming to *wz*-AlN, 3 to > 4 nm and 2-2.5 nm respectively.[8, 9] The stabilization of *zb*-AlN was also observed in AlN/W(001) superlattices with the AlN thicknesses less or equal to 1.5 nm.[10] Interface play important role in controlling their properties.[11-13]

Recent *ab inito* study shows that at ambient temperature and pressure, the most stable phase of AlN is the hexagonal wurtzite structure, while at high pressures and temperatures the rock-salt phase becomes more stable.[6] Zincblende phase was not found to be stable at any combination of pressure and temperature. This rules out the possibility of temperature or pressure playing a role in the formation of zincblende AlN.[6] DFT based calculations have also been employed before to investigate the formation and stability of the (100) AlN/TiN, AlN/VN, and VN/TiN in the rock-salt structure.[14] More recently, first-principles molecular dynamics have been used to investigate the thermal and mechanical stability of the (100) Ti/AlN rock-salt structure, with one or two monolayers of AlN as interfacial layers.[15]

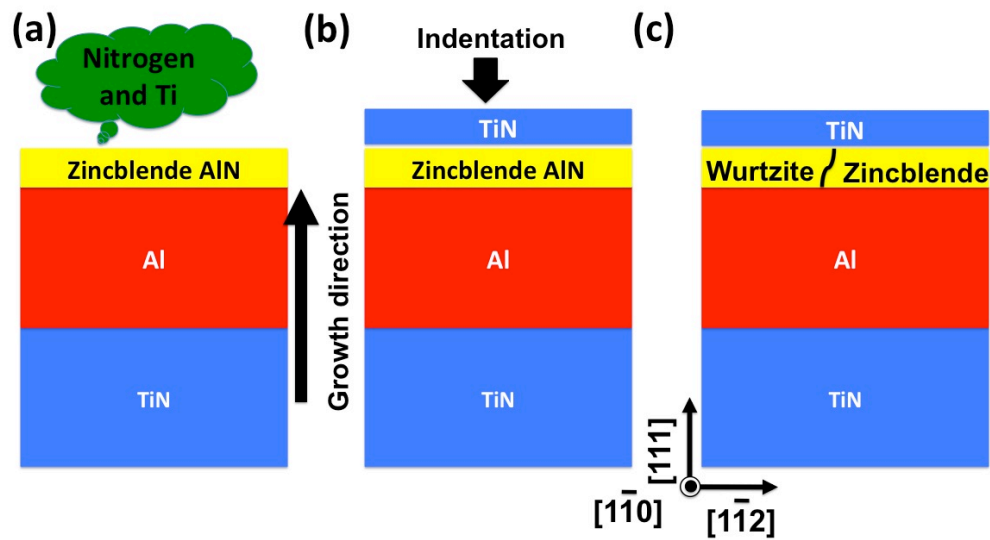

Figure 1. The schematic of Al-AlN-TiN trilayer growth. (a) Al layer is exposed to nitrogen and titanium. (b) and (c): This leads to the formation of *zb*-AlN and TiN layer. Under uniaxial indentation *zb*-AlN partially transforms to *wz*-AlN.[16]

In recent experimental work, using reactive direct current magnetron sputtering, thin multilayers of alternating Al-AlN-TiN layers were deposited at room temperature.[16] The diffraction pattern confirms the orientation relation between the Al and TiN layers, $(111)_{Al}\|(111)_{TiN}\|$interface and $<110>_{Al}\|<110>_{TiN}$, and the growth direction is along [111]. The AlN layer is about 1.5 nm thick and adopts *zb*-AlN.[16] After the indentation testing, the *zb*-AlN layer gradually transforms into the *wz*-AlN layer.[17] In figure 1(a)-(c), schematics of growth process and indentation are shown. The most intriguing findings are the growth of metastable *zb*-AlN layers only on top of Al layer and the *zb*-AlN to *wz*-AlN transformation after mechanical loading. In this paper, details of DFT calculations are presented, providing a theoretical basis to understand the energetics involved in the formation of *zb*-AlN layer, as well as for phase transformation during mechanical loading.

The rest of this paper is organized as following. The computational methods along with descriptions of the structural models are provided in Sec. 2. For presentation of DFT results in Sec. 3, we started with exploring the possibility of stabilizing AlN in zincblende phase due to various types of strains introduced during growth. Then we considered the interfaces between Al/AlN and AlN/TiN as possible sources that may thermodynamically stabilize the zincblende phase. Finally, we demonstrated that the kinetically driven diffusion of nitrogen is the only feasible explanation of the zincblende AlN formation.

## 2. Methods

Our DFT calculations were performed using the Vienna *Ab initio* Simulation Package (VASP). [18, 19] The DFT calculations employed the Perdew, Burke, and Ernzerhof (PBE) [20] generalized gradient approximation (GGA) exchange-correlation functional and the projector-augmented wave (PAW) method. [21] For all calculations, a plane wave cutoff of 500 eV for the plane wave expansion of the wave functions was used to obtain highly accurate forces. 12x12x12, 7x7x7, and 3x3x3 Monkhorst-pack mesh for k-point sampling are required to calculate the elastic constants of Al, TiN and AlN, respectively. Table 1 lists the DFT calculated values and available experimental values of lattice parameters, bulk modulus of Al, and AlN in

*wz*-AlN, *zb*-AlN, and *rs*-AlN phases, and TiN in rock-salt crystal structure. The agreement between the DFT values and the experimental data is excellent. [22, 23] The relative energies of various phases of AlN per formula unit are also listed. From Table I, it is shown that the wurtzite phase is most energetically stable, while the zincblende phase is 43 meV higher than the wurtzite phase, and the rocksalt phase is 345 meV higher.

Table 1. Comparison of calculated and available experimental values of lattice parameters, bulk modulus, of Al, TiN, and AlN in wurtzite (*wz*), zincblende (*zb*), and rock-salt (*rs*) phases.

|  | Al | | *wz*-AlN | | *zb*-AlN | *rs*-AlN | TiN | |
|---|---|---|---|---|---|---|---|---|
|  | DFT | Exp. | DFT | Exp. | DFT | DFT | DFT | Exp. |
| Lattice parameter (Å) | 4.04 | 4.05 [24] | 3.12 (a) 5.01 (c) | 3.11 (a) [25] 4.98 (c) [25] | 4.40 | 4.06 | 4.24 | 4.24 [26] |
| Bulk modulus (GPa) | 76 | 79 [27] | 202 | 211 [28] 208 [29] 220 [30] | 224 | 298 | 306 | 318 [31] |
| Relative energy (meV/f.u.) |  |  | 0.0 |  | 43 | 345 |  |  |

Interface calculations involve supercells of slabs periodically repeating in the interface plane and a vacuum region of more than 12 Å normal to the interface plane to avoid surface-surface interactions. A dipole correction perpendicular to the interface is added if both terminating surfaces are not metallic.[32] A 7x7x1 Monkhorst-Pack mesh for k-point sampling is used for all calculations involving interfaces, with 1 k-point along the largest length in the simulations. In-plane lattice parameters corresponding to the equilibrium lattice parameter of TiN and *zb*-AlN are considered. Similar methodology has been applied to our earlier metal/nitride interface calculations. [22, 33, 34]

## 3. Results and Discussions

### 3.1 Various phases of bulk AlN under stress

During growth and later under indentations, AlN may experience various types of stress. As AlN have interfaces with TiN and Al in the (111) plane so we considered three types of stress that best represent the structural constraints: 1) hydrostatic stress, 2) biaxial stress in (111) plane, 3) uniaxial stress along [111] direction.

For hydrostatic stress effect, we calculated the relative formation energy as a function of volume. Figure 2(a) shows the dependence of the ground state energies on the crystal volume. The curves correspond to the wurtzite, zincblende, and rocksalt structures. For all ranges of hydrostatic stress considered, the wurtzite phase is lower in energy compared to the zincblende phase. Under large compressive stress, the wurtzite phase transforms to rock-salt phase, which was also predicted by previous calculations. [6] The *wz*-AlN to *rs*-AlN phase transformation is accompanied by a volume reduction of 20.5% which is in good agreement with experimental values of 20.6% [35] and 18% [3] and theoretical values of 19% [4] and 22.5 [36].

As AlN interfaces with both Al and TiN, it experiences biaxial stress in the (111) plane. Fig. 2 (b) shows the relative formation energy (with zincblende as reference) per formula unit as a function of lattice parameter $a_i$. To make lattice parameter consistent for all phases, $a_i$ is *defined as* the distance between two nearest Al atoms, as shown in Fig. 3. For wurtzite crystal structure, $a_i$ equals to the hexagonal lattice parameter $a_0$. For the *wz*-AlN and *zb*-AlN, $a_i$ equals to $a_0/\sqrt{2}$, where $a_0$ is the cubic lattice constant. As biaxial strain is introduced in the (111) plane, the lattice normal to the plane is allowed to relax so that there is no stress in [111] direction. As the in-plane lattice parameter of *wz*-AlN and *zb*-AlN become smaller, the relative energy between the two compounds decreases. However, at no point *zb*-AlN becomes more stable than *wz*-AlN. In addition, at in-plane lattice parameter $a_i$ of 2.9 Å, *rs*-AlN becomes more stable than both *wz*-AlN and *zb*-AlN (see Fig. 2(b)).

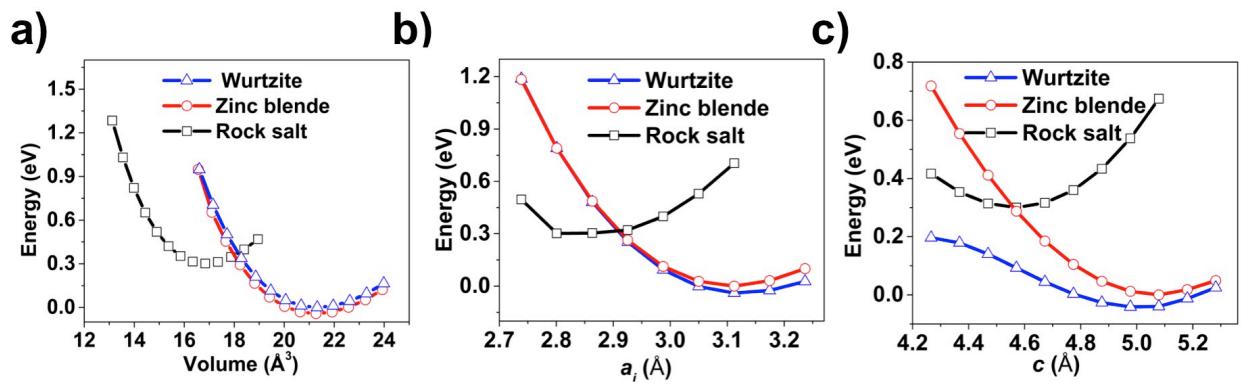

Fig. 2. Relative formation energy (zinc blende phase as reference) per formula unit under (a) hydrostatic stress as a function of crystal volume, (b) biaxial stress as a function of lattice parameter $a_i$, and (c) uniaxial stress as function of lattice parameter $c$.

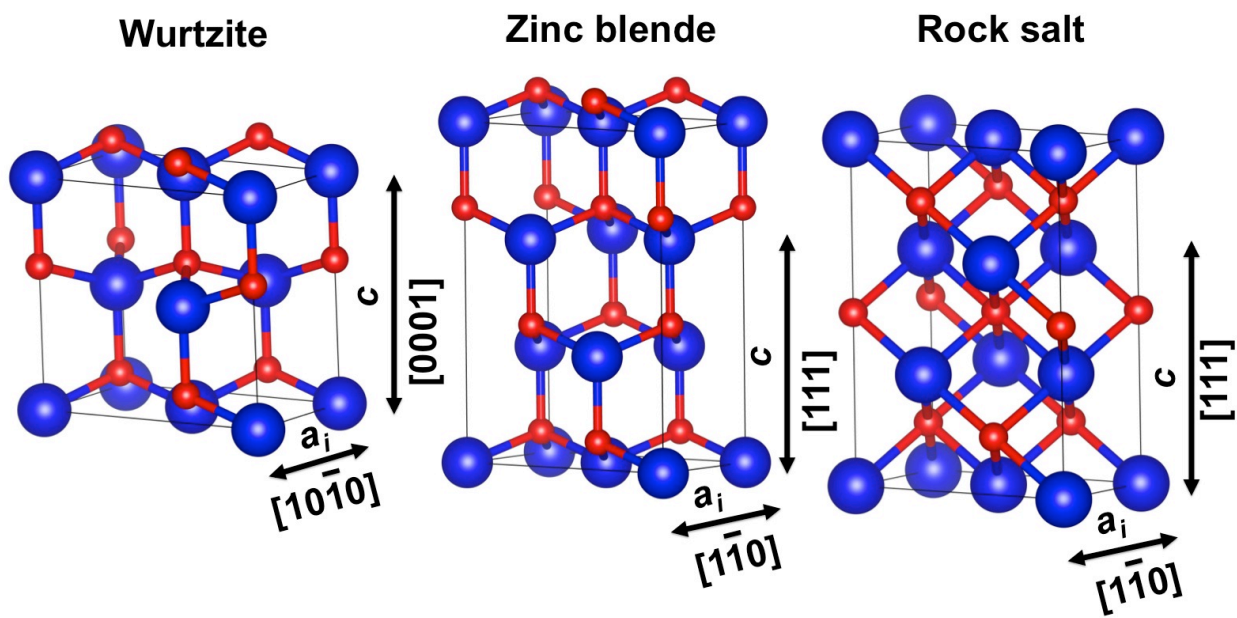

Fig. 3. Supercells of various crystal structures and definition of lattice parameters $a_i$ and $c$.

When the Al/AlN/TiN stack is indented in the [111] direction, AlN may experience uniaxial stress in the [111] direction. We calculated the relative formation energy as a function of lattice parameter $c$. To make comparison between different crystal structures easy, we again define lattice parameter $c$ for zincblende and rock-salt as 2/3 of $a_{[111]}$ or $2\sqrt{3}/3 a_0$ where $a_0$ is the cubic lattice constant, as shown in Fig. 3. For wurtzite crystal structure, $c$ equals to the hexagonal lattice parameter $c_0$ in [0001]. This is justified, as $c$ represents two stacks (AB) of Al atoms from ABABAB stacking in wurtzite phase or ABCABCABC stacking in zincblende and rocksalt phases.

Under uniaxial stresses, as the lattice is strained along the [111] direction, the lattice parameters in the (111) plane are allowed to relax so that there is no stress in the same plane. As shown in Fig. 2(c), under uniaxial compressive stresses, the difference in energy between wurtzite phase and zincblende phase becomes larger as the stress increases, with wurtzite phase more stable than zincblende phase. This explains the tendency of zincblende phase transforming to wurtzite under uniaxial compressive stress. However, under uniaxial compressive stress considered, the energy of rock-salt phase is always higher than that of wurtzite phase.

Various types of stress that might arise during growth do not explain the stabilization of zincblende phase. Next, we explore the possibility of zincblende phase stabilization due to interfaces in Al/AlN/TiN multilayers.

### 3.2 Energetics of Al/AlN and AlN/TiN bilayers

We started by exploring the relative stability of Al/AlN and AlN/TiN interfaces when AlN is in zincblende versus wurtzite phases. In-plane lattice parameters can take several values. Here for all calculations we have constrained the in-plane lattice parameters to that of *zb*-AlN. It is worth pointing out that there are several different configurations of interfaces due to the different terminations of *wz*-AlN and *zb*-AlN. We will first discuss these different terminations. In normal growth conditions, nitrogen atoms are expected to be the terminating species at the interface, as suggested in our earlier works.[33] Fig. 4(a) shows ABABA stacking of *wz*-AlN and one dangling bond (*1db*) and three dangling bond (*3db*) termination of the nitrogen layer of AlN. Similarly, Fig. 4(b) shows ABCAB stacking of *zb*-AlN and two possible nitrogen terminations of AlN: *1db* and *3db*. Thus AlN interfaced with either Al or TiN can form several types of interfaces.

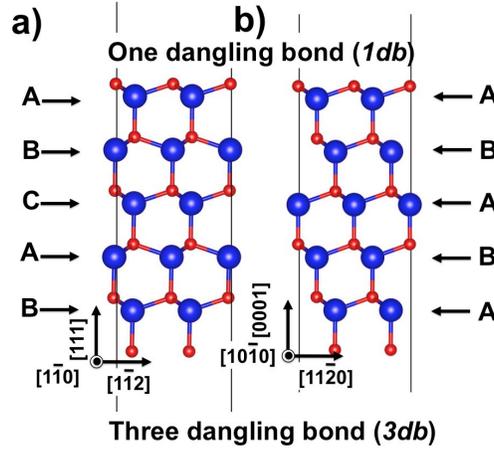

Fig. 4 Atomic structure of stacking sequence and termination of (a) left, *wz*-AlN and (b) right, *zb*-AlN.

We first studied various possible interfaces between Al/AlN. Here we only discuss the interfaces with the most stable interface structures found in terms of interface shifts relative to the bicrystals. As mentioned earlier, there are two possible terminations of AlN in both phases, with the nitrogen layer terminating in one dangling bond *1db* and three dangling bond *3db*. As shown in figure 5(a) and (b), we note that the stacking sequence of AlN for both *wz*-AlN and *zb*-AlN is the same near the interface (AB). Moreover, nitrogen atoms are located at the natural positions of each lattice type involved. As it is not possible to calculate the absolute free surface or interface energy of AlN, we calculated the relative interface energy when AlN is in zincblende verses in wurtzite phases. We assumed that the surface energies of the terminating surfaces of *wz*-AlN and *zb*-AlN are the same, which has been practiced by earlier work.[37] To calculate the relative interface energy, the total energy of the system ($E_{slab}$) is partitioned into bulk energies, surface energies, and interface energies,

$$E_{int}^{Energy} = \left( E_{Slab}^{zb/wz} - nE_{Bulk}^{zb/wz} \right) / A - S^{Al} - S^{zb/wz} \qquad (1)$$

then the relative interface energy is,

$$E_{\text{int}}^{rel} = \frac{E_{slab}^{zb} - E_{slab}^{wz} - n(E_{bulk}^{zb} - E_{bulk}^{wz})}{A} \quad (2)$$

where $E_{slab}^{zb}$ and $E_{slab}^{wz}$ are the slab energies having $zb$-AlN and $wz$-AlN respectively. $E_{bulk}^{zb}$ and $E_{bulk}^{wz}$ are bulk energies of $zb$-AlN and $wz$-AlN. n is the number of formula unit of AlN and $A$ is interface area. In this work, we did not consider interface interactions. This is equivalent to assuming that the number of AlN layers does not affect the relative interface energy. We also calculated the relative formation energy as a function of number of layers of AlN, $h_n$:

$$E_f^{rel} = E_{\text{int}}^{rel} + \frac{2h_n(E_{bulk}^{zb} - E_{bulk}^{wz})}{A} \quad (3)$$

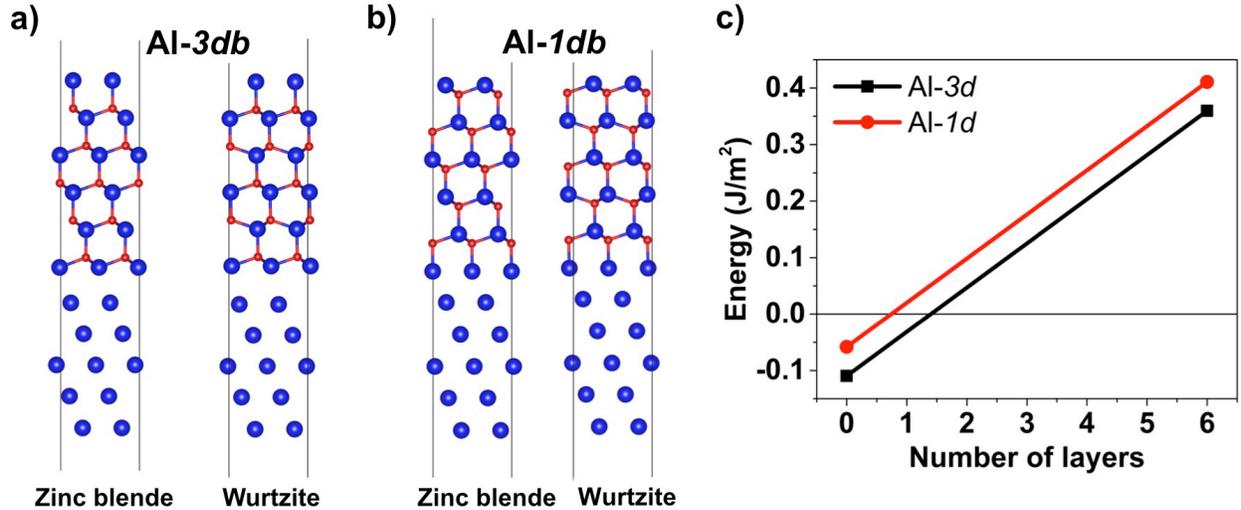

Fig. 5 Atomic structure of Al/AlN slabs (a) Al-*3db* and (b) Al-*1db*. (c) Relative stability of *zb*-AlN with respect to *wz*-AlN as a function of number of layers.

Figure 5(c) plots the excess relative energy of *zb*-AlN with respect to *wz*-AlN as a function of number of AlN layers. There are approximately 6 layers of *zb*-AlN layers during the growth in experimental conditions but interface alone only stabilizes 1 to 1.5 layers. The structure of AlN/TiN interface is similar to that of Al/TiN interface, with preference for AB stacking as shown in Fig. 6(a) and (b). Figure 6(c) plots the excess energy of *zb*-AlN with respect to *wz*-AlN as function of number of AlN layers. The interface alone only stabilizes up to two layers of *zb*-AlN.

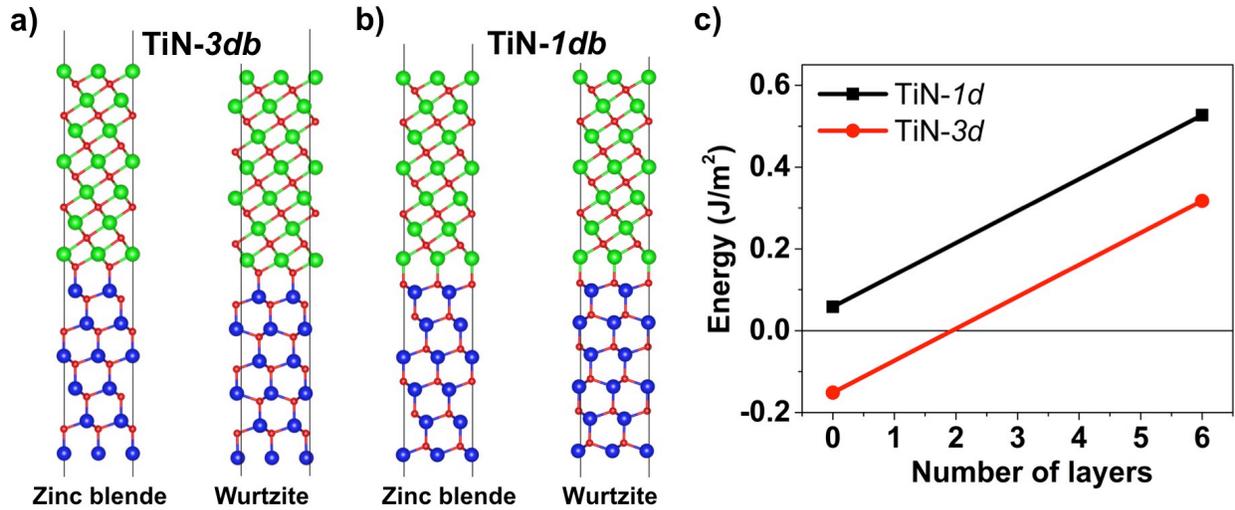

Fig. 6 Atomic structure of AlN/TiN slabs (a) TiN-*3db* and (b) TiN-*1db*. (c) Relative stability of *zb*-AlN with respect to *wz*-AlN as a function of number of layers.

### 3.3 Energetics of Al/AlN/TiN multilayers

In the previous subsection, we established the most stable terminations and configurations for Al/AlN interface and AlN/TiN interface. Here we calculate the relative interface energy with both Al/AlN and AlN/TiN interface in the same supercell. There are two possible combinations for *wz*-AlN and *zb*-AlN interfaced with Al on one side and TiN on the other side as shown in Fig. 7(a) and 7(b). In the first configuration while Al forms 3-*db* bond with the N layer of AlN, TiN forms 1-*db* bond. In the second configuration while TiN forms 3-*db* bond with the N layer of AlN, Al forms 1-*db* bond. In order to calculate the relative interface formation and relative formation energies, Eqs (2) and (3) are used respectively. Fig. 7(c) shows the relative formation energies of two configurations as a function of number of AlN layers. For all configurations considered, the maximum number of *zb*-AlN layers that interfaces can stabilize is about two, which is about 0.5 nm thickness. The change in the (111) in-plane lattice parameters does not significantly change the relative formation energy.

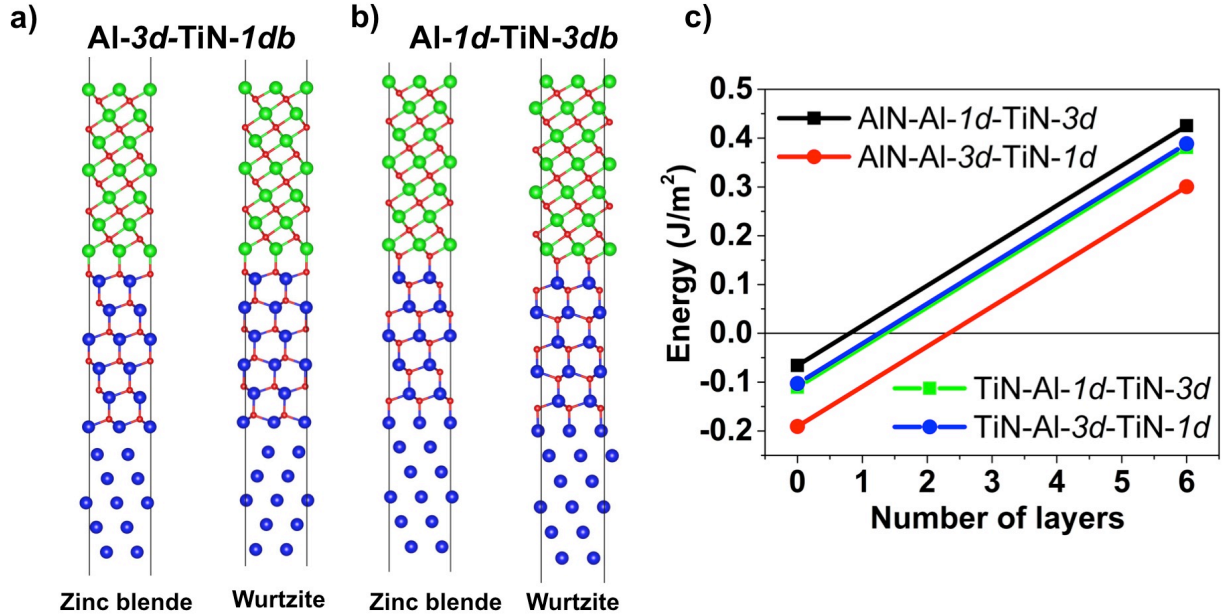

Fig. 7 Atomic structure of AlN/TiN slabs (a) Al-*1d*-TiN-*3d* and (b) Al-*3d*-TiN-*1d*. (c) Relative stability of *zb*-AlN with respect to *wz*-AlN as a function of number of layers.

Table 2. Relative interface energy calculated for various configurations at in-plane lattice parameter of TiN and *zb*-AlN. For comparison, the relative interface energy of Al/AlN and AlN/TiN and their sum (Al/AlN+AlN/TiN) are also listed.

| In-plane lattice parameter | Configurations | Al/AlN $E_{int}^{rel}\ (J/m^2)$ | AlN/TiN $E_{int}^{rel}\ (J/m^2)$ | Sum of Al/AlN and AlN/TiN $E_{int}^{rel}\ (J/m^2)$ | Al/AlN/TiN $E_{int}^{rel}\ (J/m^2)$ |
|---|---|---|---|---|---|
| TiN | Al-*1db*-TiN-*3db* | -0.210 | 0.092 | -0.118 | -0.111 |
|  | Al-*3db*-TiN-*1db* | 0.010 | -0.132 | -0.122 | -0.103 |
| *zb*-AlN | Al-*1db*-TiN-*3db* | -0.109 | 0.058 | -0.051 | -0.066 |
|  | Al-*3db*-TiN-1*db* | -0.058 | -0.151 | -0.209 | -0.182 |

Intuitively, the interface energy of Al/AlN/TiN multilayers should equal to the sum of the interface energies of bilayer Al/AlN and AlN/TiN interfaces. Table 2 summarizes results of the relative interface energy of various configurations of Al/AlN/TiN interfaces at the in-plane lattice parameters corresponding to the equilibrium lattice parameter of TiN and *zb*-AlN. Also tabulated is the relative interface energy of Al/AlN and AlN/TiN and their sum (Al/AlN+AlN/TiN). The maximum difference in the relative interface energies calculated using two methods is 0.03 J/m$^2$. This relatively small difference confirms that the assumption we made in the calculation of the relative interface energy of Al/AlN and AlN/TiN bilayers that free surface energy of *zb*-AlN and *wz*-AlN is the same, is approximately right.

**3.4 Nitridation of Al to AlN**

Now we explore the possibility of formation of *zb*-AlN by the sequential filling of N interstitials in fcc Al matrix. In Fig. 8(a), a schematic shows the Al (111) surface and various positions where nitrogen atoms can be absorbed or adsorbed. These include the tetrahedral and octahedral interstitials and on-top positions. We started by calculating the formation energy of these various positions and found that only tetrahedral and octahedral interstitial positions are stable positions.

The tetrahedral N interstitials would lead to the formation of *zb*-AlN while octahedral N interstitials would lead to the formation of *rs*-AlN. Fig. 8(b) and (c) show a sequential filling of N interstitials at (b) tetrahedral and (c) octahedral positions, up to four N atoms. In Fig. 8(d), the formation energies of these small cluster precursors to the two nitride phases are plotted as tetrahedral and octahedral positions are filled by nitrogen atoms, in two cases, one case close to the free Al(111) surfaces and the other case in bulk Al. It is clear that N interstitials occupying tetrahedral position is always energetically favorable compared to the octahedral position in these cases. We note that, at the lattice constant of Al, bulk *rs*-AlN phase actually is lower in energy than that of highly strained bulk *zb*-AlN since *rs*-AlN has a lattice constant close to that of Al. However, the preference of nitrogen atoms to sit at the tetrahedral positions in Al thus favors the formation of *zb*-AlN compared to *rs*-AlN. In earlier subsections, we already show the DFT results on the energetics of Al/AlN, AlN/TiN or Al/AlN/TiN multilayers, the experimentally observed zincblende layers can not be stabilized via interface thermodynamics. Thus, the formation of zinc-blende AlN thin film is determined to be a kinetically driven process rather than a thermodynamically driven process.

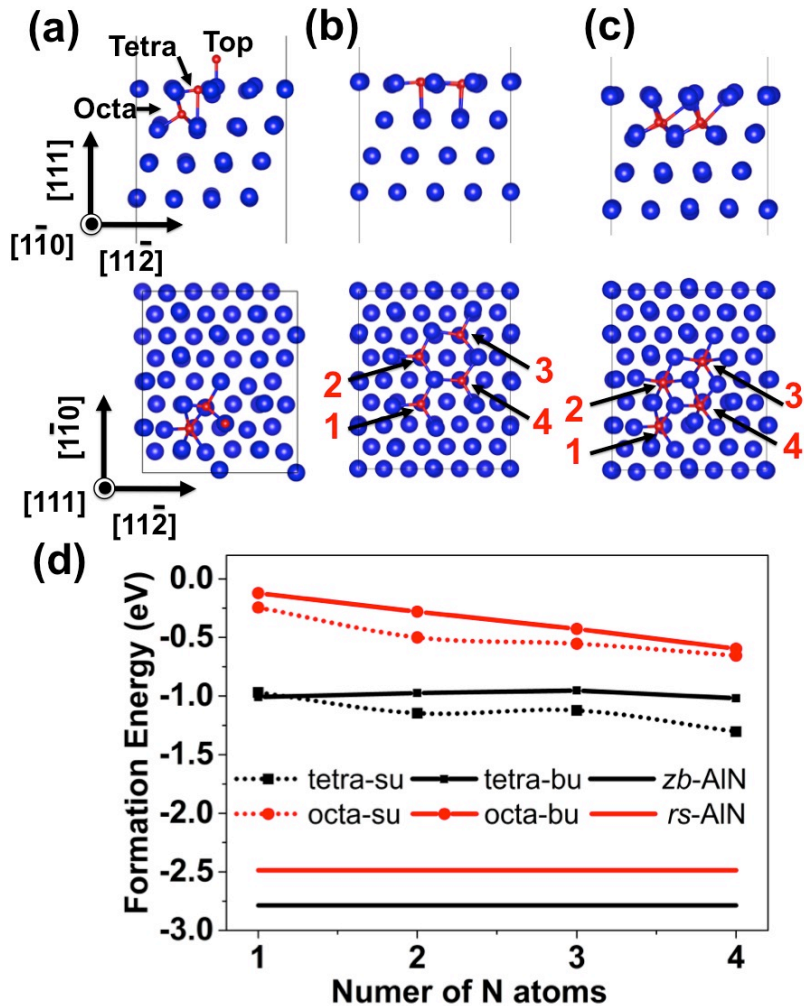

Fig. 8 (a) Various initial positions (on-top, octahedral interstitials, tetrahedral interstitials) considered when Al is exposed to nitrogen. Sequential filling of (b) tetrahedral and (c) octahedral positions, top row and bottom row are side and top views of the slab. (d) Formation energy of N interstitials per nitrogen as tetrahedral (tetra) and octahedral (octa) positions are filled close to the free surface (symbol su) and in bulk (symbol bu) calculated, and formation energy of bulk AlN in zincblende (zb-AlN) and rocksalt phases (rs-AlN).

## 4. Conclusion

In summary, using *ab initio* DFT, an unusual growth mechanism of metastable zincblende AlN thin film by diffusion of nitrogen atoms into Al lattice is established. We studied the thermodynamic stability of AlN as zincblende phase due to various epitaxial strains and found that the energy of zincblende phase is always higher than that of wurtzite phase. We then studied the relative stability of Al/AlN and AlN/TiN bilayer interfaces, and Al/AlN/TiN multilayers when AlN is in zincblende versus in wurtzite phases. For all configurations considered, the maximum number of *zb*-AlN layers that interfaces can stabilize is about two, while there are approximately six layers of *zb*-AlN layers during the growth in experimental conditions. The DFT results suggest that the interface effect by a thermodynamically driven process cannot explain the formation of zincblende AlN observed. We then compared the formation energetics of rocksalt and zincblende AlN in fcc Al through direct diffusion of nitrogen atoms to Al octahedral and tetrahedral interstitials. The preference of nitrogen atoms to sit at the tetrahedral positions in Al favors the formation of *zb*-AlN compared to *rs*-AlN. Thus, the formation of zinc-blende AlN thin film is determined to be a kinetically driven process.

**Acknowledgments**

The authors thank helpful discussions with Nan Li. This work was supported by the U.S. Department of Energy, Office of Science, Office of Basic Energy Sciences.